\documentclass[pra,aps,twocolumn,amsmath,amssymb]{revtex4}
\usepackage{graphicx}
\usepackage{dcolumn}
\usepackage{bm}
\usepackage{times}
\usepackage{epstopdf}
\usepackage{amsmath}
\usepackage{epsfig}
\newcommand{\ket}[1]{\left\vert#1\right\rangle}
\newcommand{\bra}[1]{\left\langle#1\right\vert}

\newcommand{\eq}{Eq.~}
\newcommand{\sect}{Sec.~}
\newcommand{\eqs}{Eqs.~}
\newcommand{\fig}{Fig.~}
\newcommand{\figs}{Figs.~}

\newcommand{\ie} {i.e.~}

\newcommand{\up} {\uparrow}
\newcommand{\down} {\downarrow}
\newcommand{\refs} {Refs.~}

\begin{document}
\author{G. Cordourier-Maruri\mbox{$^{1}$}}
\author{F. Ciccarello\mbox{$^{2}$}}
\author{Y. Omar\mbox{$^{3}$}}
\author{M. Zarcone\mbox{$^{2}$}}
\author{R. de Coss\mbox{$^{1}$}}
\author{S. Bose\mbox{$^{4}$}}
\affiliation{\mbox{$^{1}$}Departamento de F\'isica Aplicada, Cinvestav-M\'erida A.P. 73, M\'erida, Yucat\'an 97310, M\'exico\\
\mbox{$^{2}$}CNISM and Dipartimento di Fisica e Tecnologie Relative, Universit\`{a} degli Studi di Palermo, Viale delle Scienze, Edificio 18, 
I-90128 Palermo, Italy \\
\mbox{$^{3}$} CEMAPRE, ISEG, Universidade T\'{e}cnica de Lisboa, P-1200-781 Lisbon, and SQIG, Instituto de Telecomunica\c{c}\~oes, P-1049-001 Lisbon, Portugal\\
\mbox{$^{4}$} Department of Physics and Astronomy, University College London, Gower Street, London WC1E 6BT, United Kingdom}
\begin{abstract}

We investigate whether a two-qubit quantum gate can be implemented in a scattering process involving a flying and a static qubit. To this end, we focus on a paradigmatic setup made out of a mobile particle and a quantum impurity, whose respective spin degrees of freedom couple to each other during a one-dimensional scattering process. Once a condition for the occurrence of quantum gates is derived in terms of spin-dependent transmission coefficients, we show that this can be actually fulfilled through the insertion of an additional narrow potential barrier. An interesting observation is that under resonance conditions the above enables a gate only for isotropic Heisenberg (exchange) interactions and fails for an XY interaction. We show the existence of parameter regimes for which gates able to establish a maximum amount of entanglement can be implemented. The gates are found to be robust to variations of the optimal parameters.

\end{abstract}

\pacs{03.67.Mn, 03.67.Hk, 03.67.Lx}

\title{Implementing quantum gates through scattering between a static and a flying qubit}
\maketitle

\section{Introduction}

An emerging trend in the quest for viable ways to implement quantum information processing (QIP) tasks \cite{nc} is to envisage physical scenarios where the demanded level of control is significantly reduced. A well-known major hindrance to the reliable accomplishment of quantum coherent operations stems from the noise that any demanded manipulation of quantum ``hardware" inevitably introduces whenever a given task is to be achieved. Within this framework, an approach that is becoming increasingly popular is to encode the computational space in the (pseudo) spin degrees of freedom of scattering particles and harness their interaction during the collision to process quantum information \cite{scattering, imps,ciccarello,mappaP,mappaNP,yuasa,resil-refs,daniel,yuasaQST, teleportation}. Scattering is indeed a typical phenomenon occurring under low-control conditions: Two or more particles are prepared so as to undergo scattering and eventually measured once this has concluded. Therefore, no direct access to the very interaction process is available. Unlike gated QIP \cite{nc} where one assumes full control over interaction times to implement one- and two-qubit operations, a distinctive feature of scattering-based strategies is that any action is performed with no interaction-time tuning \cite{scattering, imps,ciccarello,mappaP,mappaNP,yuasa,resil-refs,daniel,yuasaQST,teleportation}. 
Further advantages of this approach \cite{nota1} are the remarkable resilience against a number of detrimental effects such as static disorder and imperfect setting of resonance conditions \cite{resil-refs,mappaP,daniel}, detector efficiency \cite{daniel} and decoherence affecting the centers \cite{mappaNP}. 

Evidently, the price to pay is that in quantum scattering the internal spin degrees of freedom of the scattering particles, \ie those used to encode information, inevitably couple to the motional dynamics. Hence, in general, such processes affect the state of the colliding spins according to quantum maps, instead of unitary operations. This makes the accomplishment of QIP tasks, and more in general coherent operations, rather demanding. Indeed, while the works carried out along this line targeted entanglement generation \cite{scattering, imps,ciccarello,mappaP,mappaNP,yuasa,resil-refs,daniel} and quantum state tomography \cite{yuasaQST} only latest achievements have proved the possibility to perform a quantum \emph{algorithm} such as teleportation \cite{teleportation}. The working principle behind this recent proposal, however, basically relies on a scattering-based effective projective measurement of the singlet state of two remote scattering centers \cite{teleportation}, \ie the same basic mechanism underpinning previous works that addressed entanglement generation \cite{mappaP, mappaNP, yuasa, daniel}. 

Our main motivation in the present paper is to assess whether or not a scattering-based scenario such as the one depicted above can allow for a far more ambitious task: The implementation of a two-qubit gate (TQG). Indeed, while this usually quite demanding quantum operation is well-known to be key to the settlement of a model for universal quantum computation \cite{nc}, a scattering scenario appears a hostile environment for its achievement due the above-outlined non-unitary spin dynamics.  Aware of such conditions, our main goal in this paper is to provide a proof-of-principle study to establish the possibility to implement a TQG in a simple paradigmatic model, which can serve as a milestone for forthcoming developments. To tackle the problem, we focus on a setup consisting of a quantum impurity and a mobile particle, the latter being able to propagate along a one-dimensional (1D) wire. We assume a spin-spin Heisenberg-type contact potential such that whenever the particles undergo scattering their spin degrees of freedom mutually interact. As a significant outcome, we show how basic constraints for the occurrence of quantum gates such as linearity and unitarity can be formulated in terms of spin-dependent transmission coefficients through a single, concise and physically intuitive condition. After showing that this is matched by a rather broad set of parameter patterns, we show that its fulfillment can be given a straightforward explanation in a specific regime on which we will mostly focus in this paper. 

Clearly, assuming monochromatic particles, the detrimental effect of the motional degrees of freedom in the 1D scattering process is to split the spin dynamics into a reflection and a transmission channel. Hence, similarly in some respects to other scenarios \cite{probgates} and in the spirit of the general paradigm of measurement-based quantum computation \cite{mbqc}, here our approach is to search for the occurrence of a two-channel probabilistic TQG, \ie a gate that performs one out of two given unitary operations with some associated probabilities (these are in fact reflectance and transmittance). Although our primary concern is to answer the question whether a TQG can occur regardless of its matrix form, we show that gates able to establish maximum entanglement in both the reflection and transmission channels are actually possible under certain conditions.

The present paper is organized as follows. In \sect \ref{setup} we present the aforementioned set-up and briefly discuss the approach that we adopt to describe its scattering dynamics. In \sect\ref{conditions} we derive the condition to fulfill in order for TQGs to occur. In \sect\ref{gates}, with regard to the setting introduced in \sect\ref{setup}, we illustrate the existence of a regime where the above condition holds and shed light on the explicit matrix form of the occurring gates. Finally, in \sect\ref{conclusions} we discuss the results and draw our conclusions.

\section{General setup} \label{setup}

\begin{figure}
\centerline{\includegraphics[width=0.48\textwidth]{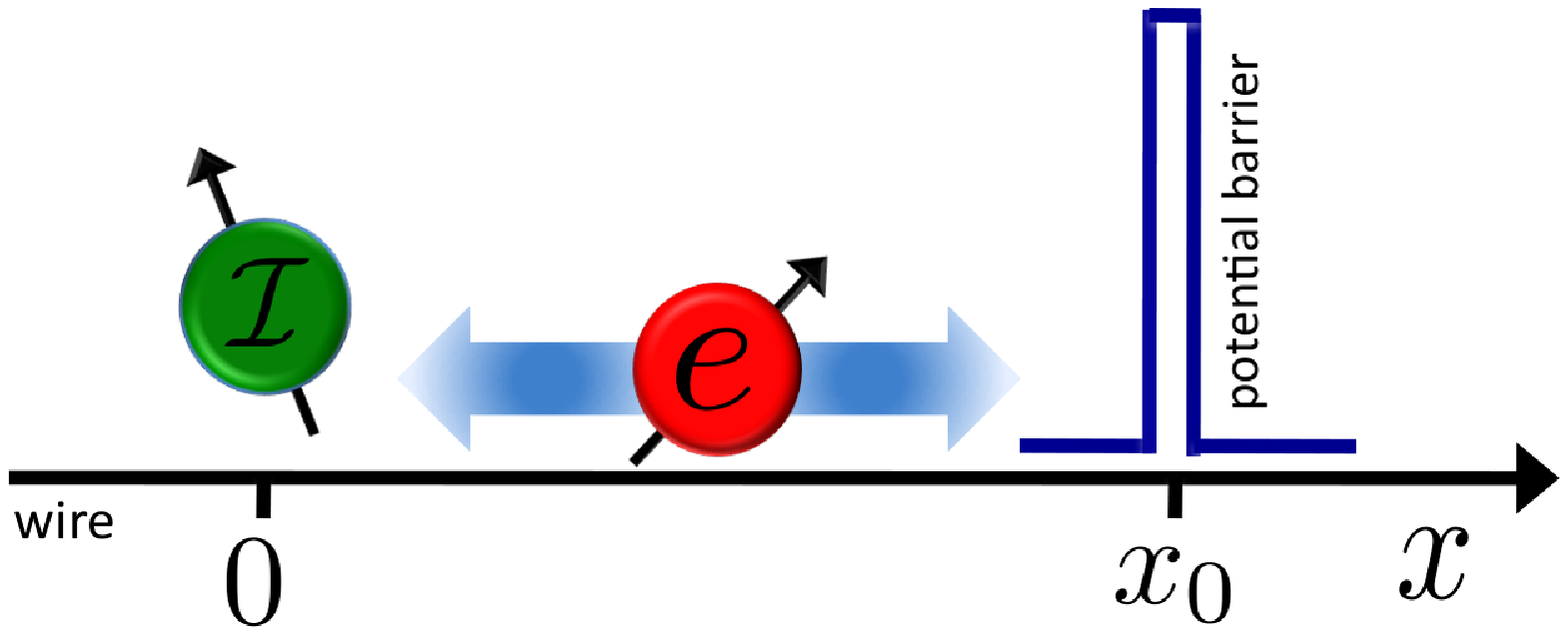}} 
\caption{(Color online) Sketch of the proposed setup for implementing the quantum gate. A mobile particle $e$ can propagate along a wire parallel to the $x$-axis. A quantum impurity $\mathcal{I}$ and a narrow potential barrier lie at $x\!=\!0$ and $x\!=\!x_0$, respectively. Once injected into the structure, $e$ undergoes multiple reflections between $\mathcal{I}$ and the potential barrier during which its spin couples to $\mathcal{I}$. Eventually, $e$ is transmitted forward or reflected back.
\label{Fig1}}
\end{figure}

We consider a 1D quantum wire along which a flying spin-1/2 particle $e$ can propagate. A quantum impurity $\mathcal{I}$, modeled as a spin-1/2 scatterer, lies at $x=0$, whereas a narrow potential barrier is located at $x=x_0$ (the $x$-axis is along the wire). The whole setting is sketched in \fig1. The Hamiltonian reads

\noindent
\begin{equation} 
\label{H}
\hat{H}=\frac{\hat{p}^{2}}{2m} + J\hat{\mbox{\boldmath$\sigma$}}\cdot\hat{\mbox{\boldmath$\mathcal{S}$}}\delta(x)+
\Gamma \delta(x-x_0),
\end{equation}

\noindent
where $m$ and $\hat{p}$ are the effective mass and momentum operator of $e$, respectively, $\hat{\mbox{\boldmath$\sigma$}}$ and $\hat{\mbox{\boldmath$\mathcal{S}$}}$ are the spin operators of $e$ and $\mathcal{I}$, respectively,  $J$ is a spin-spin coupling strength and $\Gamma$ is the potential-barrier strength (we set $\hbar\!=\!1$ throughout; notice that $J$ and $\Gamma$ have dimensions of a frequency times a length). The above paradigmatic model naturally matches within a solid-state scenarios such as a 1D quantum wire \cite{davies} or single-wall carbon nanotube \cite{devoret} with an embedded magnetic impurity or quantum dot (see also Ref.~\cite{ciccarello}). Potential barriers are routinely implemented through applied gate voltages or heterojunctions.

Clearly, due to the spin-spin contact potential [second term of Hamiltonian (\ref{H})] as $e$ enters the interaction region $0\!<\!x\!<\!x_0$ scattering along with spin flipping of $e$ and $\mathcal{I}$ take place in general. Hence, all of the scattering probability amplitudes are spin dependent. As the overall spin space is 4-dimensional, the effect of scattering is fully described by two 4$\times$4 matrices whose generic elements respectively read $t_{\alpha,\beta}$ and $r_{\alpha,\beta}$, where $t_{\alpha,\beta}$ ($r_{\alpha,\beta}$) is the probability amplitude that an initial spin state of the overall system $\ket{\alpha}_{e\mathcal{I}}$ is changed into $\ket{\beta}_{e\mathcal{I}}$ with $e$ being transmitted (reflected). Here, $\ket{\alpha}_{e\mathcal{I}}$ and $\ket{\beta}_{e\mathcal{I}}$ are two states of an orthonormal basis spanning the overall spin space.

To derive the above matrices, we first observe that according to \eq (\ref{H}) the squared total spin of $e$ and $\mathcal{I}$ as well as its projection along the $z$-axis are conserved quantities, \ie $[\hat{H},\hat{\mathbf{S}}^2]=[\hat{H},\hat{S}_z]=0$, where $\hat{\mathbf{S}}=\hat{\mbox{\boldmath$\sigma$}}\!+\!\hat{{\bf \mathcal{S}}}$ is the total spin. This entails that the dynamics within the singlet and triplet subspaces are decoupled. In each of these subspaces, the spin-spin term of $\hat{H}$ reduces to a spinless potential barrier \cite{marsiglio} so that the effective Hamiltonian describes a particle scattering from two spin-independent contact potentials as

\noindent
\begin{equation} 
\label{Heff}
\hat{H}_{S}=\frac{\hat{p}^{2}}{2m} +V_S\,\delta(x)+
\Gamma\, \delta(x-x_0)\,\,,
\end{equation} 

\noindent
where 

\noindent
\begin{equation}  \label{Vs}
V_S=\frac{J}{2} [S (S+1)-3/2]\,\,
\end{equation} 

\noindent
is an effective potential and $S$ is the quantum number associated with $\hat{\mathbf{S}}^2$ so that $S\!=\!0$ ($S\!=\!1$) in the case of the singlet (triplet). By imposing standard boundary conditions on the wavefunction and its derivative at $x\!=\!0$ and $x\!=\!x_0$ \cite{xia, marsiglio} the transmission and reflection probability amplitudes corresponding to Hamiltonian (\ref{Heff}) are straightforwardly calculated as

\noindent
\begin{eqnarray}  \label{coeffs1}
t_S&\!=\!&\frac{4}{4 + 2 i \pi \Gamma \rho_\varepsilon+ 
\pi V_S  \rho_\varepsilon \left[2 i+ (e^{2 i k x_0}\!-\!1) \pi \Gamma \rho_\varepsilon\right]}\,\,,\\
r_S&\!=\!&\pi\rho_\varepsilon\!\left\{V_S  \left[\pi\Gamma  \rho_\varepsilon \!-\!2i\right]\!-\!\Gamma e^{2i k x_0}\left[ \pi V_S  \rho_\varepsilon\!+\!2i\right]\right\}\!\frac{t_S}{4},\,\,\,\,\label{coeffs2}
       \end{eqnarray}

\noindent
where $\rho_\varepsilon\!=\!(\sqrt{2m/\varepsilon})/\pi\hbar$ is the density of states per unit length \cite{davies} of the wire, which is a function of the $e$'s kinetic energy $\varepsilon\!=\!k^2/(2m)$ ($k$ is the wavevector of $e$). It is straightforward to check that regardless of $S$ the normalization condition is fulfilled, namely

\noindent
\begin{equation}\label{norm}
|t_S|^2\!+\!|r_S|^2\!=\!1\,\,.
\end{equation}

In the above calculations, we have in fact used the coupled basis, namely the common eigenstates of $\hat{\mathbf{S}}^2$ and $\hat{S}_z$, $\mathcal{B}\!=\!\left\{\ket{\Psi^{-}}_{e\mathcal{I}},\ket{\uparrow\uparrow}_{e\mathcal{I}},\ket{\Psi^{+}}_{e\mathcal{I}},\ket{\downarrow\downarrow}_{e\mathcal{I}}\right\}$, where $|m_e\!=\up,\down\rangle_e$ ($|m_{\mathcal{I}}\!=\up,\down\rangle_{\mathcal{I}}$) are eigenstates of $\hat{\sigma}_z$ ($\hat{\mathcal{S}}_z$) and $\ket{\Psi^{\pm}}_{e\mathcal{I}}\!=\!(\ket{\up\down}_{e\mathcal{I}}\pm\ket{\down\up}_{e\mathcal{I}})/\sqrt{2}$ (henceforth we omit the particle subscripts). As the singlet and triplet spin subspaces are respectively spanned by $\ket{\Psi^{-}}$  and $\left\{\ket{\uparrow\uparrow},\ket{\Psi^{+}},\ket{\downarrow\downarrow}\right\}$ and the coefficients in \eqs (\ref{coeffs1}) and (\ref{coeffs2}) depend only on $S$ in the coupled basis $\mathcal{B}$, the transmission and the reflection probability amplitude matrices take a diagonal form

\noindent
\begin{equation} \label{T}
 \mathbf{T}\! =\!\left( \begin{array}{cccc}
  t_0&           0         &              0      &    0 \\
            0          &t_1&       0    &    0 \\
            0          &  0    &      t_1   &    0 \\
            0          &   0         &            0        &  t_1 \end{array} \right)\,\,
 \end{equation}

\noindent
(an analogous expression with the transmission coefficients replaced with the reflection ones holds for the reflection-probability-amplitude matrix $\mathbf{R}$).

\section{Condition for the occurrence of quantum gates} \label{conditions}

In general, due to the coupling between the spin and motional degrees of freedom during scattering, neither $\mathbf{T}$ nor $\mathbf{R}$ represent a unitary operator within the overall spin space. Rather, they are the matrix representations of two Kraus operators, which because of the normalization condition (\ref{norm}) fulfill 

\noindent
\begin{eqnarray}
\mathbf{T}\,\mathbf{T}^{\dagger}+\mathbf{R}\,\mathbf{R}^{\dagger}=\openone_{4}\,\,,
\end{eqnarray}

\noindent
where $\openone_{4}$ is the 4$\times$4 identity matrix. Concerning the transmission channel, the initial spin density matrix $\rho$ transforms into $\rho'$ after scattering according to

\noindent
\begin{eqnarray} \label{nonlinear}
\rho'\!=\!\frac{\mathbf{T}\rho\,\mathbf{T}^{\dagger}}{P_t}\,\,,
\end{eqnarray}

\noindent
where

\noindent
\begin{eqnarray} \label{Pt}
P_t\!=\!\mathrm{Tr}\,[\mathbf{T}\rho\,\mathbf{T}^{\dagger}] 
\end{eqnarray}

\noindent
is the overall transmission probability (analogous equations hold for the reflection channel). Because of the denominator (\ref{Pt}) \eq(\ref{nonlinear}) brings about that $\mathbf{T}$ and $\mathbf{R}$, in general, act in a nonlinear way, which would rule out the fulfillment of another essential requirement for a gate, namely linearity.
One may however wonder whether there exists a \emph{specific} regime in the setup described in \sect\ref{setup} such that unitarity and linearity occur together, a circumstance that we will show to actually take place. To this aim we consider the explicit form of the transmittivity $P_t$ in \eq (\ref{Pt}), which reads

\noindent
\begin{equation}\label{prob}
P_t=|t_0|^2 \rho_{-}\!+\!|t_1|^2 (\rho_{\up\up}\!+\! \rho_{+}\!+\! \rho_{\down\down})\,\,,
\end{equation}

\noindent
where $\rho_{\alpha}\!=\!\langle\alpha|\rho|\alpha\rangle$ ($\alpha=\up\up,\down\down$)  and $\rho_{\pm}\!=\!\langle\Psi^{\pm}|\rho|\Psi^{\pm}\rangle$ (an analogous expression clearly holds for the reflection channel). It is now straightforward to see that due to normalization of the initial spin state, \ie $\mathrm{Tr}\rho\!=\!1$, when 

\noindent
\begin{equation} \label{maincond}
|t_0|\!=\!|t_1|,
\end{equation}

\noindent
$P_t$ does not depend on $\rho$ so as to make the map (\ref{nonlinear}) linear. Furthermore, using \eq(\ref{norm}) we see that in such a case $|r_0|\!=\!|r_1|$ is fulfilled as well. In the above regime,  once rescaled operators are defined as $\mathbf{\tilde{T}}\!=\!\mathbf{T}/|t_0|$ and $\mathbf{\tilde{R}}\!=\!\mathbf{R}/|r_0|$, the spin state evolves in the transmission (reflection) channel according to $\rho'\!=\!\mathbf{\tilde{T}}\rho\mathbf{\tilde{T}}^{\dagger}$ ($\rho'\!=\!\mathbf{\tilde{R}}\rho\mathbf{\tilde{R}}^{\dagger}$). It is immediate to check that 

\noindent
\begin{eqnarray}
\mathbf{\tilde{T}}\,\mathbf{\tilde{T}}^{\dagger}=\mathbf{\tilde{T}}^{\dagger}\,\mathbf{\tilde{T}}=\mathbf{\tilde{R}}\,\mathbf{\tilde{R}}^{\dagger}=\mathbf{\tilde{R}}^{\dagger}\,\mathbf{\tilde{R}}=\openone_{4}\,\,,
\end{eqnarray}

\noindent
\ie both $\mathbf{\tilde{T}}$ and $\mathbf{\tilde{R}}$ are unitary. 

Summarizing, we have found that the simple condition (\ref{maincond}) is enough to ensure both linearity and unitarity of the transmission and reflection channels. This has indeed a reasonable interpretation due to the implicit requirement that, clearly, in order to implement a quantum gate a mere path measurement over $e$ must provide zero information about the overall spin state of $e$ and $\mathcal{I}$. It is also clear that this takes place provided that for each state of a given basis, say $\mathcal{B}$, the mobile particle is transmitted (reflected) with the same probability (and hence so happens for any spin state), a circumstance which looking at \eq(\ref{T}) we see to be equivalent to condition (\ref{maincond}).

We are now in a position to justify why our setup in \fig\ref{Fig1} includes a static potential barrier $\Gamma \delta(x-x_0)$ [see \sect \ref{setup} and \eq (\ref{H})]. In the absence of this, \ie when $\Gamma\!=\!0$,  \eq(\ref{Heff}) shows that within the singlet (triplet) subspace the effective Hamiltonian would be that of a particle $e$ scattering from a single delta-like potential barrier $-3/4 J\delta(x)$ [$J/4\,\delta(x)$]. For a particle scattered by a potential $V\delta(x)$ a straightforward textbook calculation yields that the transmission and reflection probability amplitudes $t(V)$ and $r(V)$ are given by 

\noindent
\begin{equation} \label{1barrier}
t(V)\!=\!1-r(V)\!=\!\frac{4}{4\!+\!2i \pi \rho_\varepsilon V}\,\,,
\end{equation} 

\noindent
a result which can also be obtained as a special case of \eq (\ref{coeffs1}). It is now clear that without additional scatterers there is no way to fulfill \eq(\ref{maincond}) since $|t(-3/4J)|\!\neq\!|t(J/4)|$ $\forall J$, which means that no quantum gate is possible with the simple setting consisting of $e$ and $\mathcal{I}$. In the next section, we will clearly elucidate the mechanism through which the static barrier compensates for this drawback.

We conclude this section by showing what class of quantum gates can be implemented by virtue of \eq (\ref{maincond}). As this requires that the transmittivity (reflectivity) needs be the same in the singlet and triplet subspaces, the only effect of scattering is to give rise to a relative phase shift. Therefore, in the coupled basis $\mathcal{B}$ the general form of the gate compatible with \eq (\ref{maincond}) reads

\noindent
\begin{equation}   \label{generalgate}
 \mathbf{\tilde{T}}\! =\!\left( \begin{array}{cccc}
  e^{i\varphi_t}&           0         &              0      &    0 \\
            0          &1&       0    &    0 \\
            0          &  0    &      1   &    0 \\
            0          &   0         &            0        &  1 \end{array} \right)\,\,,
\end{equation}

\noindent
where $\varphi_t\!=\!\mathrm{Arg}\, t_0\!-\!\mathrm{Arg}\, t_1$. In the computational basis $\mathcal{B}'\!=\!\left\{\ket{\uparrow\uparrow},\ket{\uparrow\downarrow},\ket{\downarrow\uparrow}, \ket{\downarrow\downarrow}\right\}$ (we encode the two qubit logical states $|0\rangle$ and $|1\rangle$ into $\ket{\up}$ and $\ket{\down}$, respectively) the gate has the general matrix representation

\noindent
\begin{equation}   \label{generalgate2}
 \mathbf{\tilde{T}}'\! =\!\left( \begin{array}{cccc}
  1&           0         &              0      &    0 \\
            0          &\frac{1\!+\!e^{i\varphi_t}}{2}&       \frac{1\!-\!e^{i\varphi_t}}{2}    &    0 \\
            0          &  \frac{1\!-\!e^{i\varphi_t}}{2}    &      \frac{1\!+\!e^{i\varphi_t}}{2}   &    0 \\
            0          &   0         &            0        &  1 \end{array} \right)\,\,.
\end{equation}

\noindent
Analogous arguments hold for the reflection channel. Later on we discuss the entangling power of the class of gates (\ref{generalgate2}).

\section{Implementing the quantum gate} \label{gates}

\begin{figure}
\centerline{\includegraphics[width=0.3\textwidth]{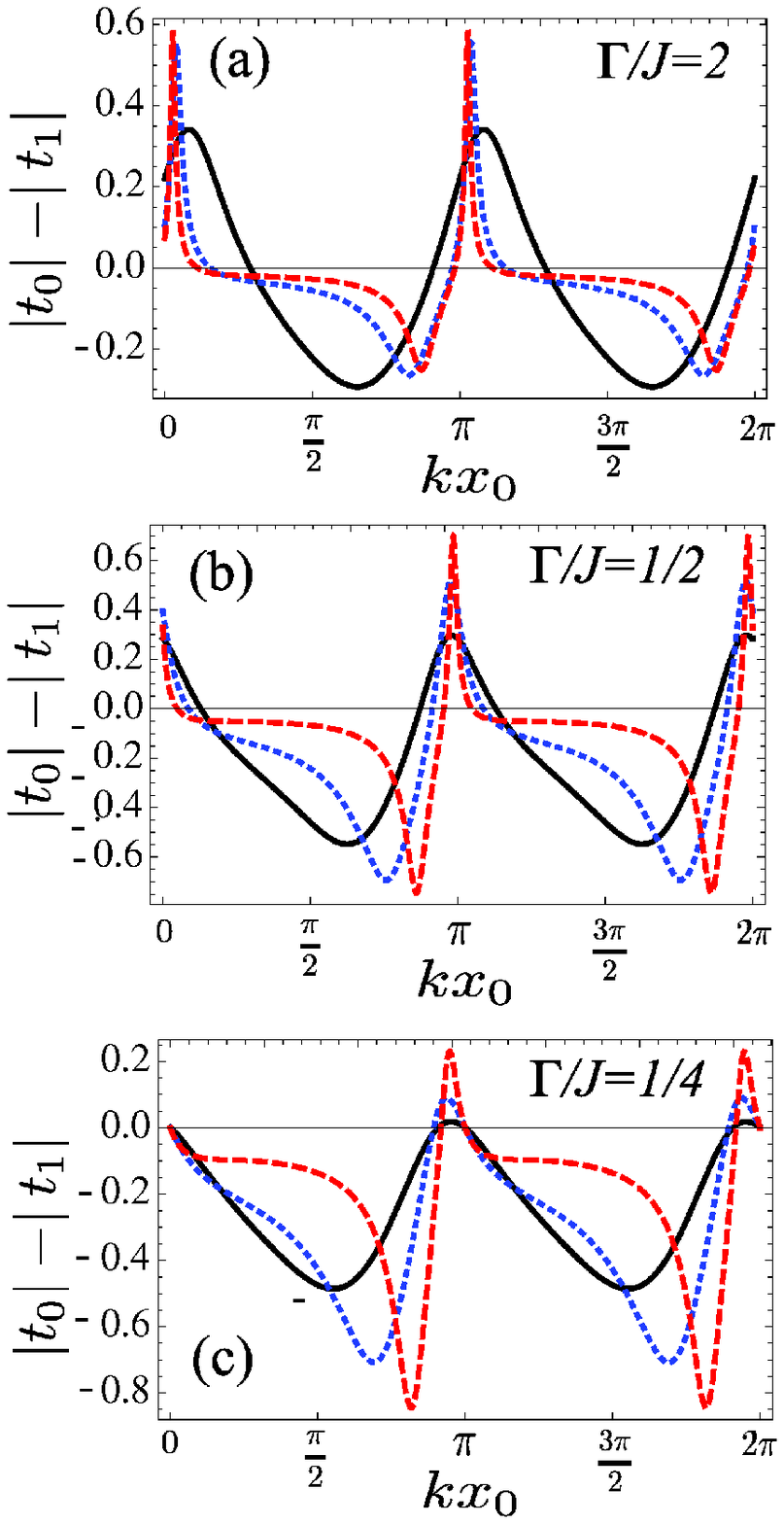}} 
\caption{(Color online) $|t_0|\!-\!|t_1|$ vs. $k x_0$ for various settings of $\Gamma/J$ and  $\rho_\varepsilon J$. (\textbf{a}) $\Gamma/J\!=\!2$ and  $\rho_\varepsilon \Gamma\!=\!0.5$ (black solid line), 2 (blue dotted) and 3 (red dashed).  (\textbf{b}) $\Gamma/J\!=\!1/2$ and  $\rho_\varepsilon \Gamma\!=\!1$ (black solid), 2 (blue dotted) and 4 (red dashed).  (\textbf{c}) $\Gamma/J\!=\!1/4$ and  $\rho_\varepsilon \Gamma\!=\!1$ (black solid), 2 (blue dotted) and 4 (red dashed).  \label{Fig2}}
\end{figure}

We now show the existence of parameter patterns such that the setup in \fig\ref{Fig1} behaves so as to satisfy  
\eq(\ref{maincond}). We recall that according to Hamiltonian (\ref{H}) the system dynamics depends on the three dimensionless parameters $\rho_\varepsilon J$, $\rho_\varepsilon \Gamma$ and $k x_0$ (see \sect\ref{setup}). In \fig\ref{Fig2}(a)-\ref{Fig2}(c), we set three different ratios between $\Gamma$ and $J$. For each of these, we plot $|t_0|\!-\!|t_1|$ against $k x_0$ for different values of $\rho_\varepsilon J$.
First, notice that for assigned values of $\Gamma/J$ and $\rho_{\varepsilon}J$ the plots are periodic in $k x_0$ with period $\pi$, in agreement with \eqs(\ref{coeffs1}) and (\ref{coeffs2}). As is evident, the condition (\ref{maincond}) occurs in each case addressed in \fig\ref{Fig2} whenever the plotted curves intersect the $k x_0$-axis. Remarkably, when $\Gamma/J\!=\!1/4$ [see \fig\ref{Fig2}(c)] provided that $k x_0\!=\!n \pi$ ($n\!\in\!\mathbb{N}$) the condition for the occurrence of quantum gates is satisfied regardless of $\rho_{\varepsilon}J$, which is of course an attractive feature for the demand of low control. Insight into this phenomenon can be given through a simple reasoning as follows. 

Under the resonance conditions (RCs) $k x_0\!=\!n \pi$ the effective representations of the two Dirac delta functions appearing in \eq(\ref{H}) coincide according to 

\noindent
\begin{equation} \label{rc}
\delta_{RC}(x)\!=\! \delta_{RC}(x\!-\!x_0)\,\,,
\end{equation}

\noindent 
where the subscripts remind that these are effective forms under RCs (for a proof see \refs\cite{ciccarello, mappaNP}, where an analogous effect has been shown to be useful for QIP tasks). Hence, in the light of \eq(\ref{rc}) under RCs the system behaves as if the static potential lay at the $\mathcal{I}$'s site. In such a case, using \eqs(\ref{Heff}) and (\ref{Vs}) the effective potentials for $S\!=\!0,1$ become $(\Gamma\!-\!3/4J)\delta_{RC}(x)$ and $(\Gamma\!+\!J/4)\delta_{RC}(x)$, respectively. As we have already shown (see previous section) when $\Gamma\!=\!0$ such single-barrier potentials necessarily entail that $|t_0|\!\neq\!|t_1|$ according to \eq(\ref{1barrier}). When $\Gamma\!\neq\!0$, however, a closer inspection at \eq(\ref{1barrier}) shows that  for a single delta-like barrier $V \delta(x)$ the modulus of the transmission coefficient only depends on $|V|$. Hence, \eq(\ref{maincond}) is fulfilled when

\noindent
\begin{equation}\label{maincond2}
\Gamma\!-\!3/4J\!=\!-(\Gamma\!+\!J/4)\,\,,
\end{equation}

\noindent
which is indeed equivalent to $\Gamma/J\!=\!1/4$ regardless of $\rho_{\varepsilon}J$, thus explaining the aforementioned behavior in \fig\ref{Fig2}(c). A more explicit and illustrative way to see this is noticing that under RCs the effective Hamiltonian can be arranged as

\noindent
\begin{equation} 
\label{otra}
\hat{H}=\frac{\hat{p}^{2}}{2m} + J\left[\frac{\Gamma}{J}+\hat{\sigma}_z\hat{S}_z + \frac{\hat{\sigma}_+\hat{S}_- +\hat{\sigma}_-\hat{S}_+}{2}\right]\delta_{RC}(x)\,\,.
\end{equation} 

\noindent
When the initial spin state is $\ket{\up\up}$ or $\ket{\down\down}$ the factor between brackets takes the value $\Gamma/J+1/4$, which results in the effective potential-barrier height $\Gamma+J/4$. On the other hand, $\ket{\Psi^{\pm}}$ fulfill

\noindent
\begin{eqnarray}
\hat{\sigma}_z\hat{S}_z \,\ket{\Psi^{\pm}}&=&-\frac{1}{4} \ket{\Psi^{\pm}}\,\,,\\
\frac{\hat{\sigma}_+\hat{S}_- +\hat{\sigma}_-\hat{S}_+}{2}\,\ket{\Psi^{\pm}}&=&\pm\frac{1}{2} \ket{\Psi^{\pm}}\label{pot}\,\,.
\end{eqnarray}

\noindent
It is now immediate to see that the static barrier, whose presence is embodied by the constant term $\Gamma/J$ between the brackets in \eq(\ref{otra}), in fact cancels out the Ising term for $\Gamma/J\!=\!1/4$. When this takes place the effective potential-barrier height seen by $\ket{\Psi^{\pm}}$ becomes $\pm J/2$, whose modulus is the same as the one associated with $\ket{\up\up}$ and $\ket{\down\down}$. The above reasoning also shows, in particular, that the replacement of a Heisenberg-type spin-spin coupling with an $XY$-isotropic one in Hamiltonian (\ref{H}) cannot give rise to any quantum gate either with no extra barriers or with a $\delta$-like barrier under RCs. Indeed, in the Heisenberg case one deals with only two independent transmission coefficients  ($t_0$ and $t_1$) and hence the single condition (\ref{maincond}) needs to be fulfilled to implement gates. In the $XY$-isotropic model, however, {\it three} independent coefficients in general arise since the spin-spin scattering potential vanishes for both $|{\up\up}\rangle$ and $|{\down\down}\rangle$ and  takes the effective value $\pm J/2$ for $|\Psi^{\pm}\rangle$ [see \eq(\ref{pot})]. In the light of our discussion in \sect\ref{conditions}, we see that in the case of $XY$-isotropic model gates' occurrence results in two equations to be fulfilled. Absence of gates under RCs with such a model is therefore not surprising given that setting of RCs in fact freezes the parameter $k x_0$: judicious setting of the remaining free parameter $\Gamma$ is enough to fulfill the single equation required by the Heisenberg model, but not the two ones met with $XY$-isotropic coupling. 

\begin{figure*}
\centerline{\includegraphics[width=0.9\textwidth]{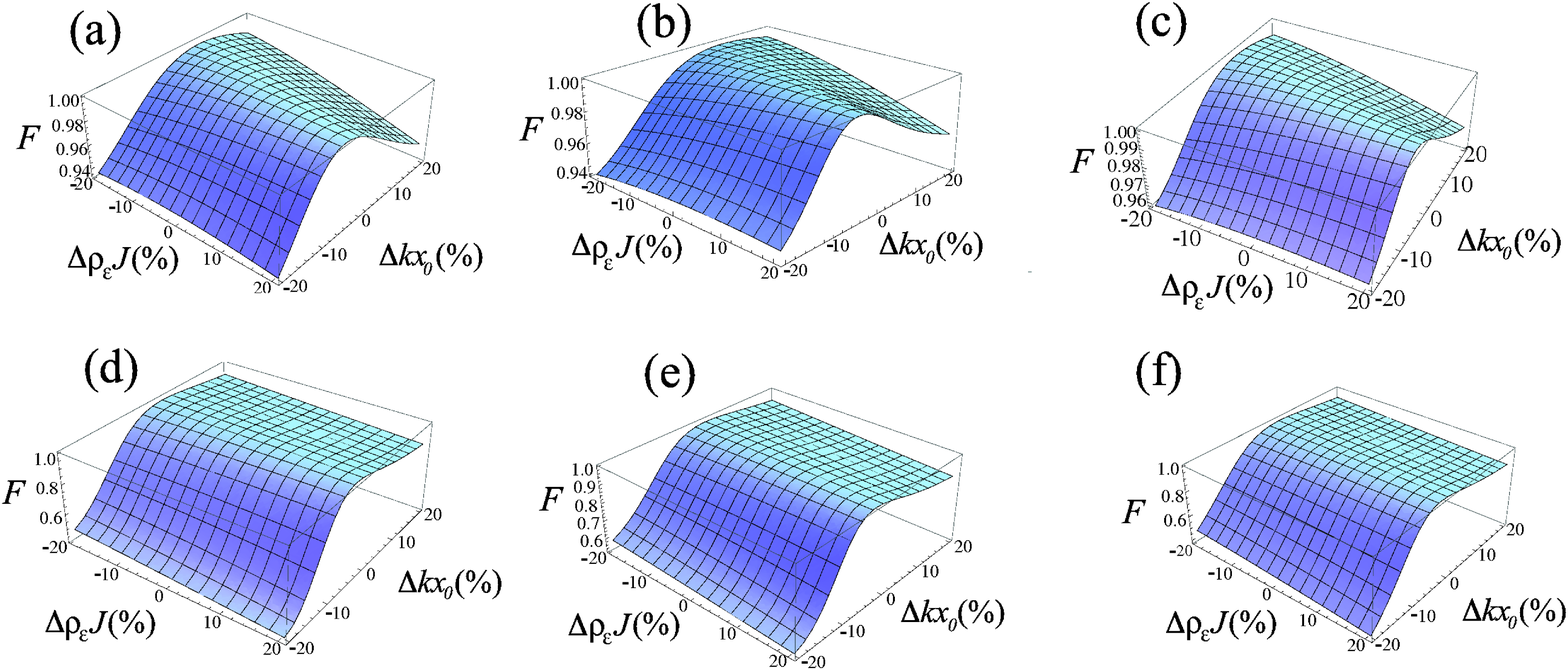}} 
\caption{(Color online) Fidelity $F$ vs.~the percentage deviations from the ideal values $\Delta \rho_\varepsilon J$ and $\Delta k x_0$ for the initial spin states $\ket{\up\down}$ [(\textbf{a}) and (\textbf{d})], $(\ket{\up\up}\!+\!\ket{\down\down}\!+\!\ket{\up\down}\!-\!\ket{\down\up})/2$ [(\textbf{b}) and (\textbf{e})] and $(\ket{\up}\!+\!\ket{\down})\bigotimes(\ket{\up}\!+i\!\ket{\down})/2$ [(\textbf{c}) and (\textbf{f})]. Plots (\textbf{a}), (\textbf{b}) and (\textbf{c}) refer to the transmission-channel gate, whereas (\textbf{d}), (\textbf{e}) and (\textbf{f}) refer to the reflection-channel one. \label{Fig3}}
\end{figure*}

Having identified a regime compatible with \eq(\ref{maincond}), our next task is to illustrate what specific forms of $\mathbf{\tilde{T}'}$ and $\mathbf{\tilde{R}'}$ can occur within the general family in \eq(\ref{generalgate2}), which is in fact equivalent to explore what values $\varphi_t$ and $\varphi_r$ can take. In the regime $\Gamma/J\!=\!1/4$ and $k x_0\!=\!n\pi$, a straightforward calculation along with use of \eqs(\ref{Vs}) and (\ref{1barrier}) yield

\noindent
\begin{eqnarray}\label{phitphir}
\varphi_t\!&=&\!2\,\arctan\,\frac{\pi\rho_\varepsilon J}{4} \label{phit}\,\,, \\
\varphi_r\!&=&\!-2\,\arctan\,\frac{4}{\pi\rho_\varepsilon J}\label{phir}\!+\!2\pi\,\,.
\end{eqnarray}

\noindent
In the light of \eq(\ref{generalgate2}), \eqs(\ref{phit}) and (\ref{phir}) fully specify the form taken by gates $\mathbf{\tilde{T}'}$ and $\mathbf{\tilde{R}'}$ in the regime $\Gamma/J\!=\!1/4$ and $k x_0\!=\!n\pi$. Thus both the phase shifts $\varphi_t$ and $\varphi_r$ grow with $\rho_\varepsilon J$ tending to an asymptotic value, which is $\pi$ in the case of $\varphi_t$ and $2\pi$ in the case of $\varphi_r$ (for any $\rho_\varepsilon J$ we have $\varphi_t\!-\!\varphi_r\!=\!\pi$).

A question that is naturally raised from the matrix structure in \eq(\ref{generalgate2}) is whether the elements of the central 2$\times$2 block can all have the same modulus. Indeed, in such a case the gate is clearly able to establish \emph{maximum entanglement}. It is immediate to see that the above circumstance occurs provided that $|1\!+\!e^{i\varphi_t}|\!=\!|1\!-\!e^{i\varphi_t\!}|$, which requires $\cos \varphi_t\!=\!0$ and hence $\varphi_t\!=\!(2q\!+\!1)\pi/2$, where $q\!\in\!\mathbb{Z}$ (analogous arguments hold true for the reflection channel). In our case, using \eqs(\ref{phit}) and (\ref{phir}) we obtain $\varphi_t\!=\!\varphi_r\!-\!\pi\!\!=\!\pi/2$ for $\rho_{\varepsilon}J\!=\!4/\pi\!\simeq\!1.27$, which entails the gate matrix form

\noindent
\begin{equation}   \label{generalgate3}
 \mathbf{U}\! =\!\left( \begin{array}{cccc}
  1&           0         &              0      &    0 \\
            0          &\frac{1\pm i}{2}&      \frac{1\mp i}{2}  &    0 \\
            0          & \frac{1 \mp i}{2}   &      \frac{1\pm i}{2} &    0 \\
            0          &   0         &            0        &  1 \end{array} \right)\,\,,
\end{equation}

\noindent
where the $+$ ($-$) sign holds for the transmission (reflection) channel. Also, using \eqs(\ref{Vs}) and (\ref{1barrier}) it is immediately checked that in such a case $|t_0|\!=\!|t_1|\!=\!|r_0|\!=\!|r_1|\!=\!1/2$, namely the success probabilities associated with the reflection and transmission gates are the same. We have thus found a parameter pattern such that gates able to create maximum entanglement occur in both the reflection and transmission channels. Taking for instance the initial product state $\ket{\up\down}$ we obtain that, up to an irrelevant phase factor, $\hat{U}\ket{\up\down}\!=\!(\ket{\up\down}\!\mp i\ket{\down\up})/\sqrt{2}$ where the $-$ ($+$) sign holds for the transmission (reflection) channel.

As we have proven, in order for our set-up to implement gates, certain parameter values need to be set. This feature may appear somewhat unnatural in the low-control scattering scenario that we have considered. A legitimate question is therefore how robust is the gate to an imperfect setting of the ideal parameters. To answer this, we consider the paradigmatic situation where one wishes to implement the maximally entangling gate (\ref{generalgate3}) discussed above, which requires to set $\Gamma/J\!=\!1/4$, $\rho_{\varepsilon}J\!=\!4/\pi$ and $k x_0\!=\!n\pi$. To measure how well such gate is implemented for an imperfect matching of this ideal pattern we use quantum fidelity \cite{nc}. Specifically, for a given initial pure spin state $\rho\!=\!\ket{\Psi_i}\!\bra{\Psi_i}$ we compute the fidelity $F$ between the output state $\rho'$ as given in \eq(\ref{nonlinear}) (in general this is mixed) and the output state $\ket{\Psi_f}\!=\!\hat{U}\ket{\Psi_i}$ that would be obtained in the ideal case. The expression of fidelity is $F\!=\!\bra{\Psi_f}\rho'\ket{\Psi_f}$. In \fig3, we have carried out this study for the transmission and reflection channels [\figs3(a)-(c) and \figs3(d)-(f), respectively] and the three representative initial states $\ket{\up\down}$, $(\ket{\up\up}\!+\!\ket{\down\down}\!+\!\ket{\up\down}\!-\!\ket{\down\up})/2$ and $(\ket{\up}\!+\!\ket{\down})\bigotimes(\ket{\up}\!+i\!\ket{\down})/2$. 
In each case, we set the condition $\Gamma/J\!=\!1/4$ and plot $F$ against $\Delta \rho_\varepsilon J$ and $\Delta k x_0$, where $\Delta \rho_\varepsilon J$ ($\Delta k x_0$) is the percentage difference between $\rho_\varepsilon J$ ($k x_0$) and the corresponding ideal value. As is evident in the plots, the robustness of the transmission-channel gate is quite striking. For deviations from the ideal values up to 20$\%$, in the worst case $F$ slightly decreases to $\simeq\!0.94$. On the other hand, the reflection channel exhibits generally lower performances \cite{resilience} especially for negative values of $\Delta k x_0$.  In this channel, however, for $|\Delta k x_0|$ up to $\simeq 8\%$ $F$ exceeds $0.9$. 

Such generally good resilience is in line with the outcomes of analogous tests in similar set-ups \cite{resil-refs,mappaP,daniel}, which further confirms a major advantage of scattering-based methods to accomplish QIP tasks (see Introduction). 

\section{Conclusions} \label{conclusions}

In this work we have tackled the issue whether a two-qubit gate (TQG) can be implemented in a set-up made out of mobile and static qubits undergoing quantum scattering processes. Despite the many advantages of such a scattering scenario for QIP purposes \cite{scattering, imps,ciccarello,mappaP,mappaNP,yuasa,resil-refs,daniel,yuasaQST, teleportation} and the well-known importance of TQGs \cite{nc}, this question had so far remained fully unanswered in the literature. With these motivations in mind, we have considered a minimal paradigmatic set-up comprising a flying spin scattering from a quantum impurity along with a further spinless potential barrier. In a way similar to other scenarios where proposals for probabilistic quantum gates were put forward \cite{probgates} we have assessed whether a unitary transformation in the overall spin space can be probabilistically implemented in each of the transmission and reflection channels. By imposing basic constraints such as linearity and unitarity, we have found that gates occur in both channels provided that a simple and physically intuitive condition is obeyed. We have also given the full class of resulting gates. Next, numerical evidence has been given that the above theoretical condition is actually matched for suitable parameter patterns in both off-resonance and resonance conditions. After focusing on RCs, we have analytically derived the exact parameter pattern that ensures the occurrence of gates. Insight into the related underlying mechanism has been given by explaining, in particular, the essential role played by the additional potential barrier. Among the possible occurring gates, we have identified one able to establish maximum entanglement and given the exact required parameter setting. Finally, we have shown that such maximally entangling gate is robust against imperfect matching of the optimal parameters.

As anticipated, a significant implication of our findings is the ability of certain occurring gates to establish maximum entanglement. Entanglement between a static and a flying qubit offers the major advantage of being particularly prone to a robust Bell test since once scattering has happened the two particles can get significantly far apart from each other.

In this work, we have mainly focused on the accomplishment of gates under RCs. Indeed, this regime is more prone to analytical treatment than the more general off-resonance case. This enabled us to highlight a number of key issues without mathematical hindrances. However, as we have seen, even off-resonance conditions allow for occurrence of TQGs. A comprehensive study of this case, which goes beyond the scopes of this paper, is therefore highly desirable and will be the subject of a future publication \cite{future}.

The entanglement between a static ionic and a flying photonic qubit has been envisaged for connecting up ion trap quantum registers \cite{dmn}. Likewise, It is quite possible that the scheme we propose here will open up scaling opportunities for spin-based quantum computation in solid state systems \cite{loss}.

\begin{acknowledgments}
Fruitful discussions with  D.~E.~Browne are gratefully acknowledged. YO would like to thank the support from project IT-QuantTel, as well as from Funda\c{c}\~{a}o para a Ci\^{e}ncia e a Tecnologia (Portugal), namely through programs POCTI/POCI/PTDC and project PTDC/EEA-TEL/103402/2008 QuantPrivTel, partially funded by FEDER (EU). GC and RC wish to thank the Department of Physics and Astronomy, UCL, and SQIG, Instituto de Telecomunica\c{c}\~oes for their hospitality as well as the financial support from CONACYT (Mexico) under Grant No. 83604.
\end{acknowledgments}


\begin{thebibliography}{18}
\bibitem{nota1} In this work, differently from those in Ref.~\cite{scattering} we focus on the scattering-based QIP paradigm which exploits scattering centers possessing internal spin degrees of freedom \cite{imps,ciccarello,mappaP,mappaNP,yuasa,resil-refs,daniel,yuasaQST,teleportation}. In the remainder of the paper, it is therefore the latter scenario that we have in mind when using the term ``scattering".
\end{thebibliography}
\end{document}